\documentclass[preprint2]{aastex}

\usepackage{graphicx}
\usepackage{xfrac}
\usepackage{enumerate}

\begin{document}

\title{PIRATE: A Remotely-Operable Telescope Facility for Research and Education}

\author{Holmes\altaffilmark{1}, S., Kolb\altaffilmark{1}, U., Haswell\altaffilmark{1}, C. A., Burwitz\altaffilmark{2}, V., Lucas\altaffilmark{1}, R. J., Rodriguez\altaffilmark{3}, J., Rolfe\altaffilmark{4}, S. M., Rostron\altaffilmark{5}, J., Barker\altaffilmark{1}, J.}
\altaffiltext{1}{Department of Physics and Astronomy, The Open University, Milton Keynes, MK7 6AA, UK}
\altaffiltext{2}{Max-Planck-Institut f\"{u}r extraterrestrische Physik, Giessenbachstrasse, 85748, Garching,  Germany}
\altaffiltext{3}{Observatori Astron\`{o}mic de Mallorca, Cam' de l'Observatori, 07144 Costitx, Mallorca, Spain}
\altaffiltext{4}{Planetary and Space Sciences Research Institute, The Open University, Milton Keynes, MK7 6AA, UK}
\altaffiltext{5}{Department of Physics and Astronomy, University of Warwick, Coventry, CV4 7AL, UK}

\begin{abstract}

We introduce PIRATE, a new remotely-operable telescope facility for use in research and education, constructed from `off-the-shelf' hardware, operated by The Open University.  We focus on the PIRATE Mark 1 operational phase where PIRATE was equipped with a widely-used 0.35m Schmidt-Cassegrain system (now replaced with a 0.425m corrected Dall Kirkham astrograph). Situated at the Observatori Astron\`{o}mic de Mallorca, PIRATE is currently used to follow up potential transiting extrasolar planet candidates produced by the SuperWASP North experiment, as well as to hunt for novae in M31 and other nearby galaxies. It is operated by a mixture of commercially available software and proprietary software developed at the Open University. We discuss problems associated with performing precision time series photometry when using a German Equatorial Mount, investigating the overall performance of such `off-the-shelf' solutions in both research and teaching applications. We conclude that PIRATE is a cost-effective research facility, and also provides exciting prospects for undergraduate astronomy. PIRATE has broken new ground in offering practical astronomy education to distance-learning students in their own homes.


\end{abstract}

\section {Introduction}
PIRATE (Physics Innovations Robotic Astronomical Telescope Explorer) is a remote telescope facility situated on the Balearic island of Mallorca ($2^{\circ} 57' 03.34'' $E, $39^{\circ} 38' 34.31''$ N), at the Observatori Astron\`{o}mic de Mallorca, $\sim$162m above sea level. It is a remotely-operable facility used for research and undergraduate teaching, allowing for a practical astrophysics component to be implemented within The Open University's (OU, Milton Keynes, UK) distance-learning modules. PIRATE is constructed entirely from `off-the-shelf' hardware, and is operated primarily with commercial observatory control software, with some additional proprietary software developed to support simultaneous use by groups of students.  The total cost of the facility, combining all purchased hardware and software is of order \$150k.

The funding for PIRATE was primarily made available by a teaching innovation initiative (piCETL \footnote{http://www8.open.ac.uk/opencetl/physics-innovations-centre-excellence-teaching-and-learning}) to explore the integration of a remotely-controlled observatory into university-level distance teaching modules. The aim of the project was to reproduce the hands-on experience of a traditional lab course with real-time access to a telescope from any computer linked to the internet.

 In this paper we refer to the PIRATE Mark 1 facility (unless mentioned otherwise) which featured a Celestron\footnote{http://www.celestron.com} C14 0.35m Schmidt-Cassegrain Optical Tube Assembly (OTA). This OTA is currently no longer in use within the PIRATE facility, having been superseded in August 2010 by a 0.425m PlaneWave Systems\footnote{http://www.planewave.com} CDK17 f/6.8 Corrected Dall-Kirkham Astrograph telescope, which makes use of a new custom micro-focuser (provided by PlaneWave),  which we designate PIRATE Mark 2. The Mark 1 hardware remains an obvious choice for cost-effective astronomy, so we report in detail on its capabilities, features, and drawbacks. 

\subsection{Use in Education}

PIRATE needs to cater for a relatively large student body with little or no prior experience in observational astronomy, and with a very limited amount of remote supervision by a tutor. Granting a sufficiently high level of access to the facility, such as the ability to initiate the opening and closing of the dome, is considered essential for the student learning experience. The adopted software solution, a combination of the commercial product ACP\footnote{http://www.dc3.com} \citep{denny} and the proprietary software described in \cite{lucas2010}, achieves the desired student access privileges without compromising the safe operation of PIRATE. After one tutor-led induction evening conducted via audio communication and web interfaces, the student groups could operate PIRATE successfully on their own. A (remote) night-duty astronomer was on call but only occasionally contacted. 

\begin{figure}[htp]
\begin{center}
\plotone{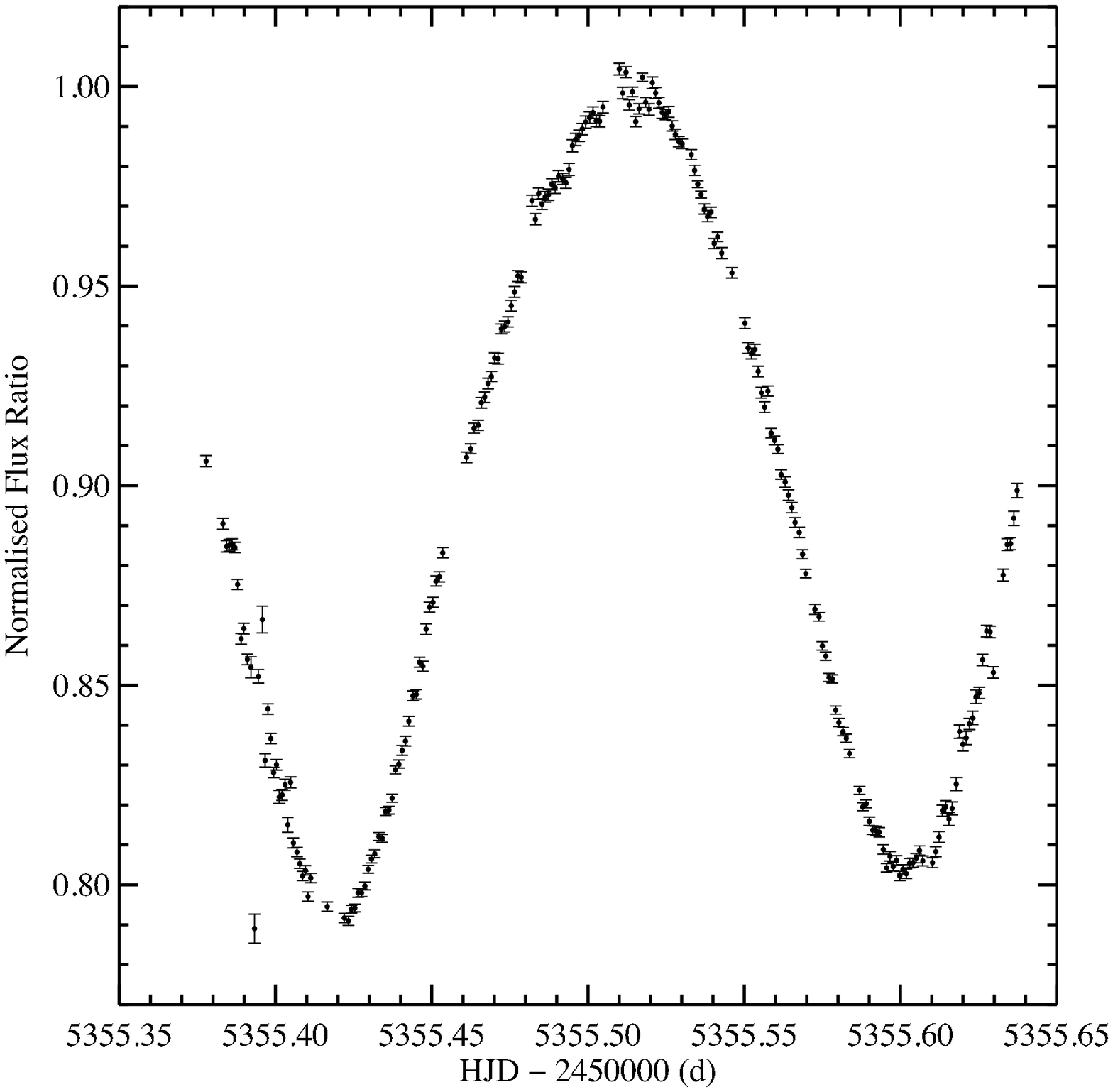}
\caption{\small{The light curve created from 206 I band images of 1SWASPJ163420.90+424433.4. The data were taken to fulfill project objectives for the OU's undergraduate course, S382, by OU undergraduate students.}}
\label{fig:eblc}
\end{center}
\end{figure}

PIRATE was deployed for OU undergraduate students for the first time in spring 2010 in a 10 week project which is part of the third level (third year) OU module S382 ÒAstrophysicsÓ. A total of about 30 students formed three groups with alternating access to PIRATE, for a total of 40 observing nights. Each group selected a suitable target source from the catalogue by \cite{norton} of periodic variables found by SuperWASP and coincident with a ROSAT source. The groups then built up a long-term light curve of their target, such as the example shown in Fig. \ref{fig:eblc}. Individual observing sessions were staffed by small observer teams of 2-4 students who kept in audio and text contact throughout the night. The collaboration of different observer teams in the larger groups on the same target source ensured the emergence of a useable data base to which all group members could develop a sense of ownership, even when occasional nights were clouded out and the observer teams on duty may not have succeeded in obtaining data themselves. A fuller account of the challenges and solutions for the PIRATE teaching project can be found in \cite{kolb2010}.

\subsection{Use in Research}

In its research role, PIRATE performs follow-up photometry of SuperWASP Transiting Extrasolar Planet (TEP) candidates from the SuperWASP (Wide Angle Search for Planets) \citep{pollacco2006} survey, and discovers extragalactic novae.

Wide-field transit surveys produce light curves of millions of objects. Transit search algorithms \citep{colliercameron2006} are applied to trend-filtered (e.g. \cite{tamuz}) light curves to select TEP candidates from the millions of surveyed objects. Most of these candidates will not be TEPs, but will be `astrophysical false positives' (AFPs), which are estimated to outnumber TEPs in a transit search survey by at least an order of magnitude \citep{odonovan2006}. High priority candidates are followed up with radial velocity measurements to confirm the mass and therefore planetary nature of the transiting object, but this requires scarce large-telescope spectroscopy.

To winnow TEP candidates from wide-field surveys like SuperWASP, small-to-medium sized telescopes provide higher precision, higher spacial resolution follow-up photometry \citep{haswell2010}. In this role, PIRATE can act as a link in the planet-finding chain, reducing the amount of large telescope time spent on false-positives.

PIRATE has also detected novae in M31 (e.g. M31N 2010-01a) \citep{burwitz2010}, as part of an XMM-Newton large program for studying the source population in the Andromeda Galaxy, M31, in which a large number of supersoft X-ray sources were identified as counterparts of optical novae \citep{pietsch2008}. This requires the long term monitoring of M31 in X-rays with XMM-Newton, Chandra, and occasionally SWIFT. PIRATE forms part of a wider network of optical telescopes that routinely monitor M31 to discover optical novae and track their light curves. Optical spectra of the freshly discovered novae are also obtained with larger telescopes, such as the Hobby-Eberly Telescope, USA, and the BTA, Russia, to classify the novae. With the help of the information extracted from these X-ray and optical data we are beginning to obtain a better picture of the physical processes that take place in nova outbursts. This optical observing campaign grew out of the XMM-Newton/Chandra M31 nova monitoring collaboration \citep{pietsch2010}.

In the following sections we describe the hardware and software used in the PIRATE facility, as well as detailing commissioning issues, data reduction techniques, and future improvements.

\section {Hardware and Software}

\begin{figure}[htp]
 \begin{center}
  \plotone{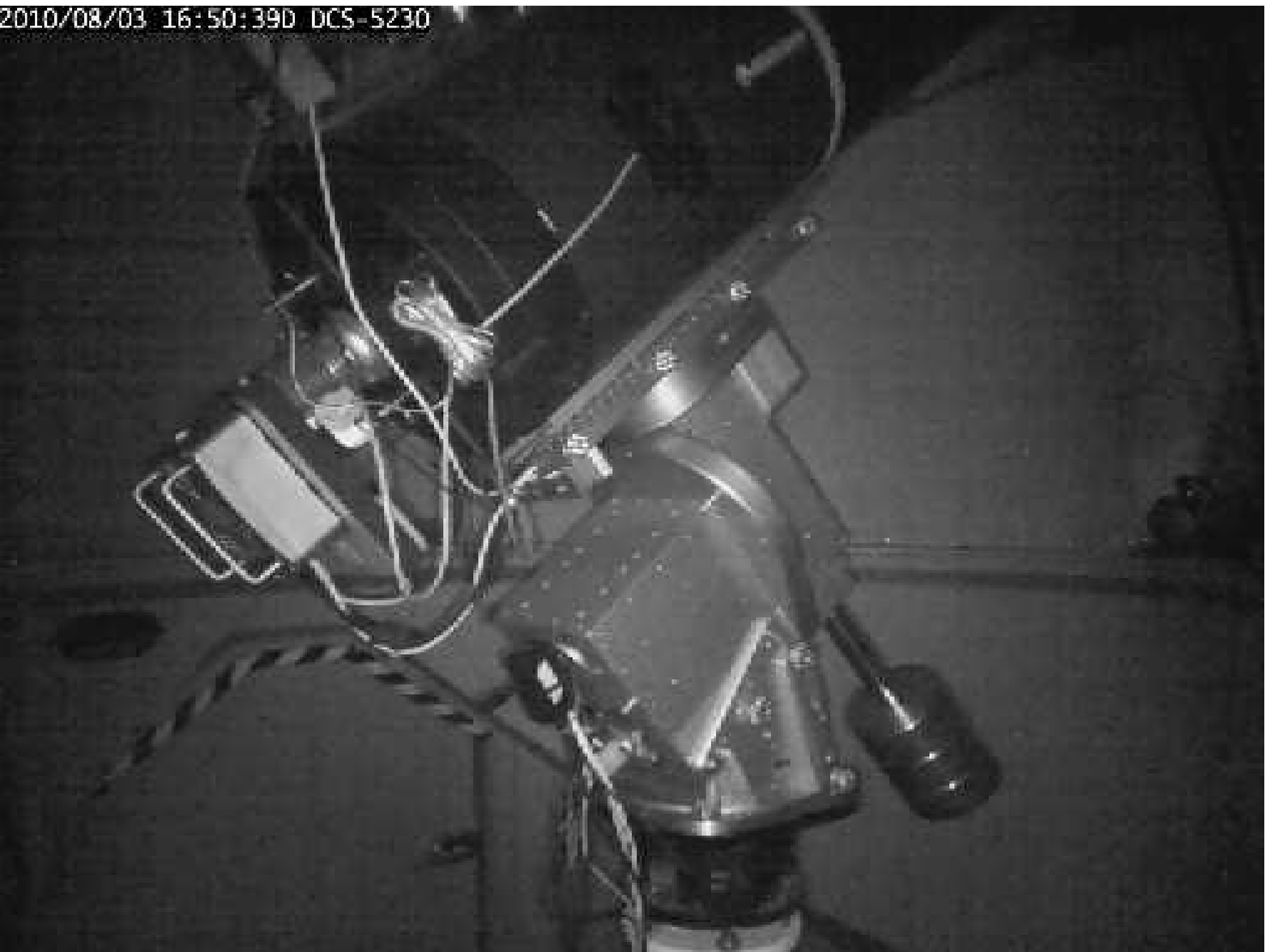}
  \caption{\small{The C14 as seen through the D-LINK DCS-5230 IR webcam}}
  \label{fig:piratephot}
  \end{center}
  \end{figure}

\begin{figure}[htp]
 \begin{center}
  \plotone{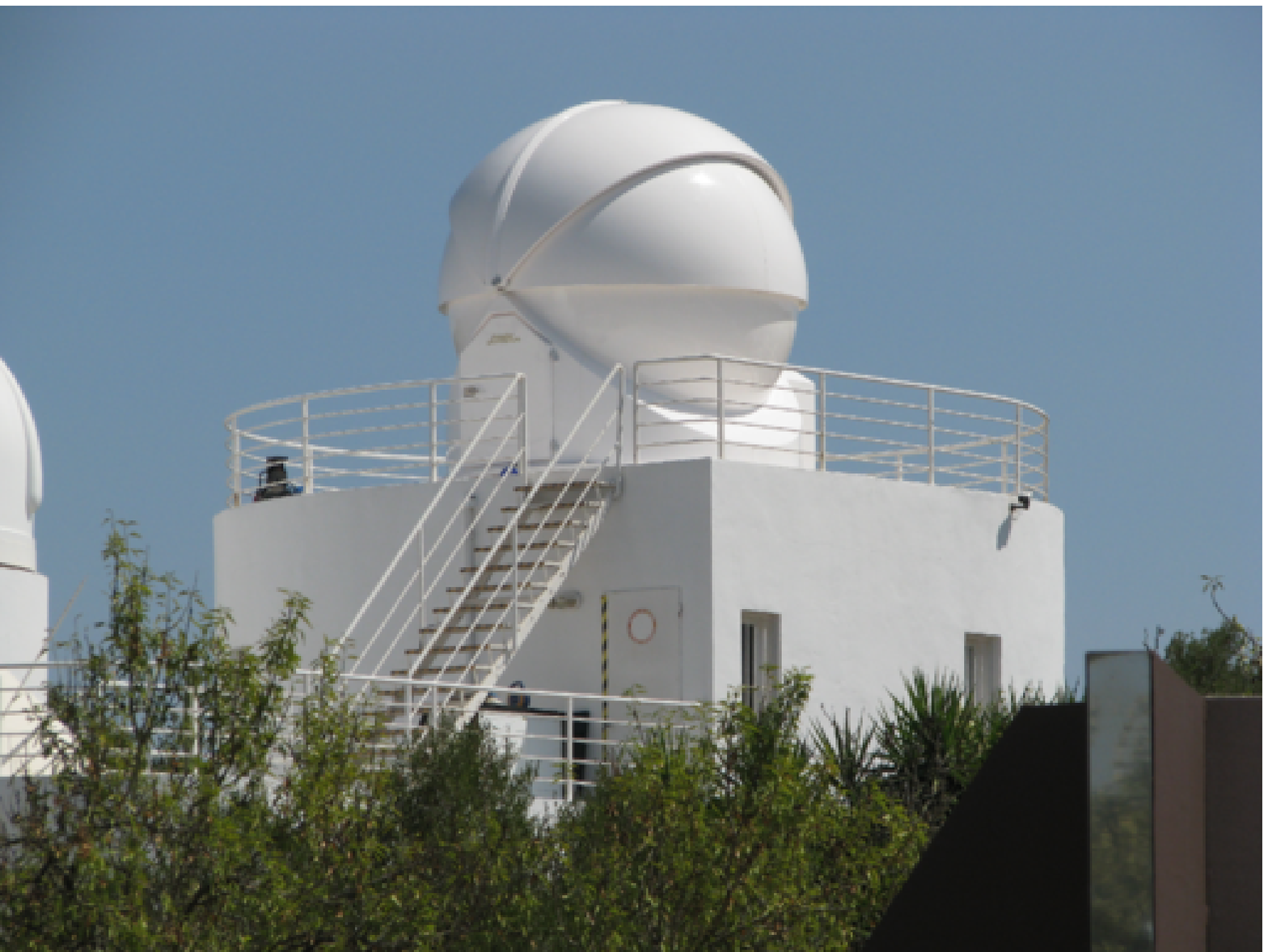}
  \caption{\small{The 3.5m Baader Planetarium All Sky dome. The control room is directly below.}}
  \label{fig:dome}
  \end{center}
  \end{figure}

PIRATE Mark 1 is a remote telescope facility consisting of a 14$''$ Schmidt-Cassegrain reflector mounted on a Paramount ME\footnote{http://www.bisque.com} German Equatorial Mount (GEM) (Fig. \ref{fig:piratephot}). Its main imager is an SBIG STL-1001E, which houses the Kodak KAF-1001E CCD, a $1024 \times 1024$ pixel front-illuminated detector with a pixel size of $24\mu$, and a quantum efficiency of 0.4, 0.55, and 0.65 for wavelengths 450, 550, 650nm respectively\footnote{KAF-1001E spec. sheet: http://www.kodak.com}. In combination with the 3.91m focal length of the Celestron C14, this produces a plate scale of 1.21$''$ pixel$^{-1}$, and a field of view of $22' \times 22'$. An Optec TCF micro-focuser is located between the photometer enclosure and optical assembly, along with an 8 position filter wheel, containing Johnson-Cousins standard BVRI broadband filters and 3 narrowband filters (H$\alpha$, SII, OII). PIRATE has a Celestron 80mm refractor with SBIG ST402 ME camera for auto-guiding. The telescope is housed within a Baader Planetarium\footnote{http://www.baader-planetarium.com/} 3.5m diameter All-Sky clamshell dome (Fig. \ref{fig:dome}) with built in weather systems for automatic shut-down in adverse conditions.  On-site weather is monitored by a weather station and Boltwood cloud sensor (to the side of the dome), and both internal and external dome-mounted weather systems (rain, internal humidity, internal temperature). The dome has its own firmware which interfaces with the proprietary dome driver, as well as with the Boltwood cloud sensor, allowing for shutdown conditions to be communicated directly to the dome, bypassing the control PCs. Four D-Link webcams (3 internal, 1 external) provide live video and audio feeds, as well as IR beams for night-viewing.

\subsection{Camera Characteristics}
\begin{figure}[htp]
 \begin{center}
  \plotone{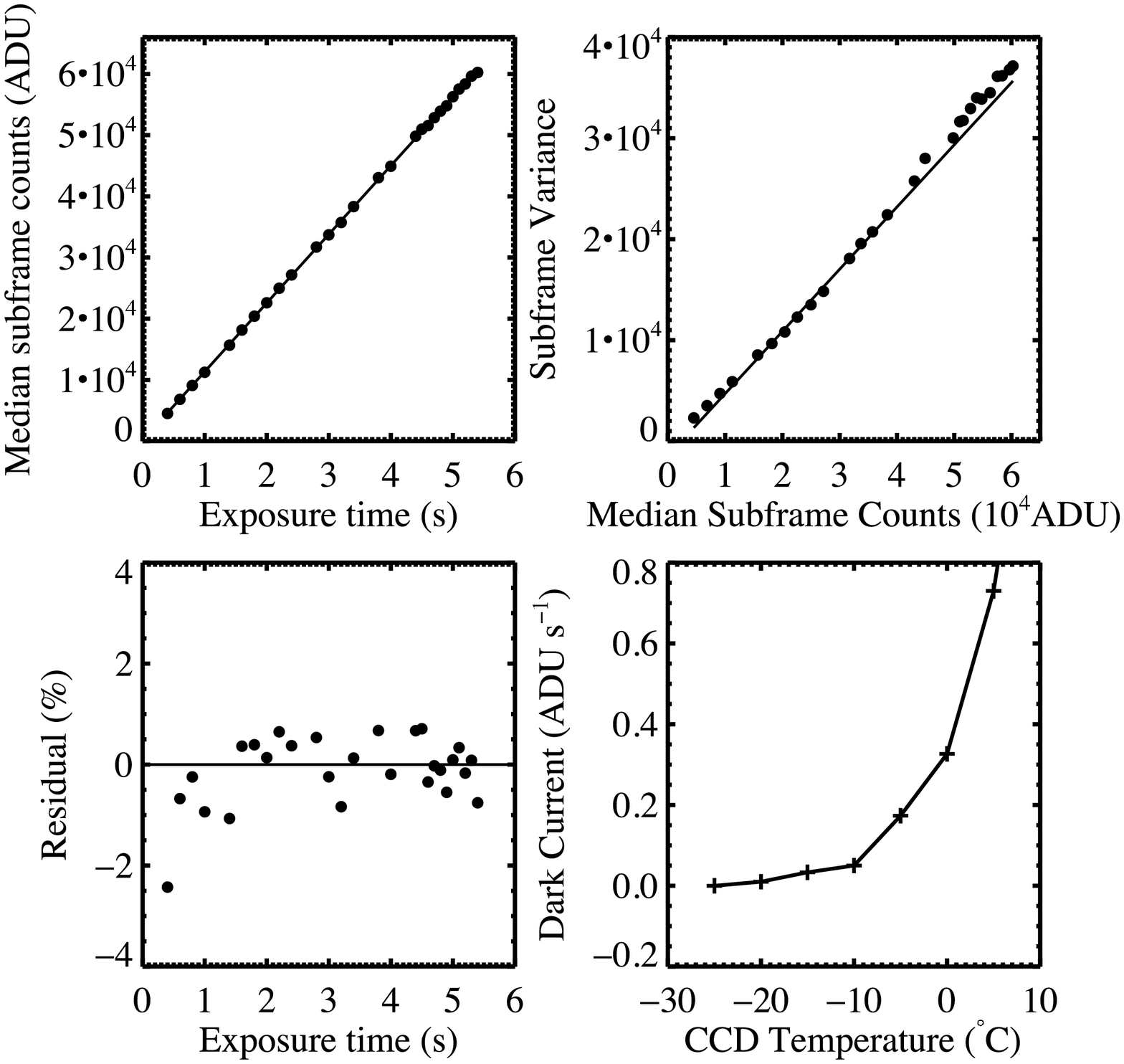}
  \caption{\small{{\bf (a) Top left:} Counts as a function of exposure time, with a linear fit through the expected linear region 20k - 40k ADU. {\bf (b) Bottom left:} the residuals of the above fit. {\bf (c) Top right:} the photon transfer curve, with linear fit to the same 20k - 40k ADU region. The deviation from linearity at $~\sim$0 - 10k ADU is due to the transition from the shot noise to the read noise regime. {\bf (d) Bottom right:} dark current as a function of chip temperature}}
  \label{fig:ccdplot}
  \end{center}
  \end{figure}

	To characterise the CCD we determined the gain, linearity, read noise and the dark current (which was measured at a range of temperatures). The gain and the linearity measurements were made using the same set of dome flats for each. A set of dome flats of increasing exposure time were taken in order to measure the median counts of a 100 $\times$ 100 pixel subframe in the middle of each image. For the short exposure times (of order a few seconds), we attempted to generate a shutter correction map using the methodology of \cite{zissell}, anticipating that we might see position-dependent corrections to the exposure time (due to the shutter travel) of order $10^{-3}$s. However, this produced a null result, and further investigation into the shutter mechanism of the STL-1001E confirms a rotating `shutter wheel' that should be devoid of shutter travel effects. We therefore do not apply any shutter correction to the short exposure flat fields. We measured the bias level from contemporaneous bias frames and subsequently subtracted the mean pedestal level of 107 ADU from each flat. The subframe was chosen to be the centre of the vignetting function, where the image is at its flattest. To assess the linearity of the CCD, we plotted median subframe counts against exposure time (Fig. \ref{fig:ccdplot} (a)), and fit a linear trend to the same ADU range. The residuals of this fit can be seen in Fig. \ref{fig:ccdplot} (b). We note a deviation from linearity at the top end of the dynamic range of $<$1\%, and $<$2\% at the bottom end. We measured the gain using the following relationship: $\sigma_{ADU}^2 = \frac{1}{g} \langle N_{ADU}\rangle$ \citep{howell}, where $\sigma_{ADU}$ is the standard deviation of the sub-frame counts, and $\langle N_{ADU}\rangle $ is the mean recorded counts in the subframe. To derive this expression, one assumes the statistical relationship $\sigma_{e^-} = \sqrt{N_{e^-}} $ holds, where $\sigma_{e^-}$ and $N_{e^-}$ are the uncertainty in and the number of recorded photoelectrons respectively, which only holds when the process is governed by Poisson statistics.  To determine $g$ we therefore plot $\sigma_{ADU}^2$ against $ \langle N_{ADU}\rangle$ (see Fig. \ref{fig:ccdplot} (c)). We needed sufficient photoelectrons for photon-counting (Poisson) noise to be dominant, and so we fitted to the range 20k ADU $< \langle N_{ADU}\rangle <$ 40k ADU. We chose an upper limit of 40k ADU to be well clear of the digital counting limit of 65535 ADU (16-bit). We measured a value for the gain of $1.62 \pm 0.01$ e$^{-}$ADU$^{-1}$. The manufacturers quote a full-well capacity of 150k e$^{-}$, which, given the determined gain, would occur at a theoretical ADU value of
$\sim$ 92.5k ADU, an unobtainable value due to the digital counting limit.  We note that falling short
of the electron full well capacity is probably responsible for our excellent linearity of response at the upper end of our digital counting range.

	To measure the read-out noise, we took 30 bias frames at a temperature of $-20^{\circ}$, and median combined them. The median-combined bias frame (which we assume to have negligible read noise) was then subtracted from each of the 30 bias frames, and the standard deviation of each of these difference images measured, we averaged this figure to achieve the read noise in ADU \citep{howell}. We measure then (using the previously determined gain value) a read-out noise of $10.9 \pm 1.3$ e$^{-}$. To check for consistency, we also split the bias frames into pairs, subtracting one from the other within each pair, so that $d= b_1 - b_2$, where $b_1$ \& $b_2$ are the bias frames in each pair, and $\sigma_{read} = \sigma_{d}/\sqrt{2}$, yielding a read-noise measurement for each pair. These values were consistent with the previous read-noise measurement.
	
	Dark current was assessed for a range of temperatures (from 15$^{\circ}$C through to -25$^{\circ}$C in increments of 5$^{\circ}$C) by median-combining 10 darks at each temperature, 10 biases at each temperature, subtracting the bias and dividing the residual counts by the exposure time (30s for each dark frame). The measured values are listed below in Table \ref{tab:darktable}. We measure a dark current of $0.02 \pm 0.02$ e$^{-}$s$^{-1}$ and $0.08 \pm 0.02$ e$^{-}$s$^{-1}$ at -20$^{\circ}$C and -10$^{\circ}$C, the typical operating temperatures for winter and summer respectively. We also find a value of $0.51\pm 0.03$ e$^{-}$s$^{-1}$ at 0$^{\circ}$C, c.f. the 1 e$^{-}$s$^{-1}$ at 0$^{\circ}$C manufacturer's specification.	

\begin{table}	
\begin{tabular}{ c c c }
 \hline  \hline
T ($^{\circ}$C) & D. C. (e$^{-}$ s$^{-1}$) & $\sigma_{D.C.}$ (e$^{-}$ ADU$^{-1}$) \\ \hline
-25 & 0.01 & 0.02 \\
-20 & 0.02 & 0.02 \\
-15 & 0.05 & 0.02 \\ 
-10 & 0.08 & 0.02 \\
 -5   & 0.27 & 0.03 \\
 0    & 0.51 & 0.03 \\
 5    & 1.13 & 0.03 \\
 10  & 2.25 & 0.03 \\
 15  & 5.20 & 0.04 \\ \hline
\end{tabular}
\caption{\small{Measured dark current for a range of chip temperatures}}
\label{tab:darktable}
\end{table}
	
\subsection{The Mount}

The Paramount ME GEM is able to deliver an all-sky pointing accuracy of around 2$'$.  We employ the software `TPoint'\footnote{http://www.tpsoft.demon.co.uk/} to model system flexure to achieve this value, in which we use 30 model points that capture the discrepancy between intended and actual pointing for a variety of telescope positions. We had previously employed a $\sim$150 point model, which was abandoned due to the addition of new hardware. We noted that the subsequent 30 point model replacement provides sufficient pointing accuracy to enable the control program to plate-solve and correct for pointing inaccuracy with only a small contribution to inter-frame overhead, and so have settled at 30 points for convenience. We also have the ability to identify and remove periodic error in the sidereal tracking from within the mount control software, by measuring the correcting impulses provided by the autoguider. The periodic error is caused by irregularities in the rotation rate of the worm gear through one cycle, and so presents itself entirely in RA. We fitted a 5\textsuperscript{th} order polynomial with a peak-to-peak amplitude of 4.8$''$, and subsequently saw improvements in the circularity of each star's point spread function. As this is a German Equatorial Mount, the OTA must `flip' from one side of the pier to the other when the tracked target crosses the meridian. This results in the observed field being rotated 180$^{\circ}$ with respect to the optics and camera.

\subsection{Software}

The hardware is controlled by a collection of Windows-based software (MaxIm DL\footnote{http://www.cyanogen.com} for imaging, TheSky6\footnote{http://www.bisque.com} for mount control, FocusMax\footnote{http://users.bsdwebsolutions.com/~larryweber/}\citep{weber} for control of the micro-focuser) which are each controlled by ACP\footnote{http://acp.dc3.com}, an observatory control suite that automates the command of each of the component software packages through the ASCOM-standard driver layer. These programs are run on a 2.13GHz CPU Windows XP PC with 2Gb RAM. The site receives its internet connection via satellite; connection speeds are good, with typical upload speeds from the OAM to the UK $\sim$ 250kBs$^{-1}$, and operating latencies of around 60-100ms. We use Dimension 4 from Thinking Man Software \footnote{http://www.thinkman.com/dimension4} to synchronise the control PC's clock to UTC via an NTP server every 10 seconds.

\begin{figure}[htp]
\begin{center}
\plotone{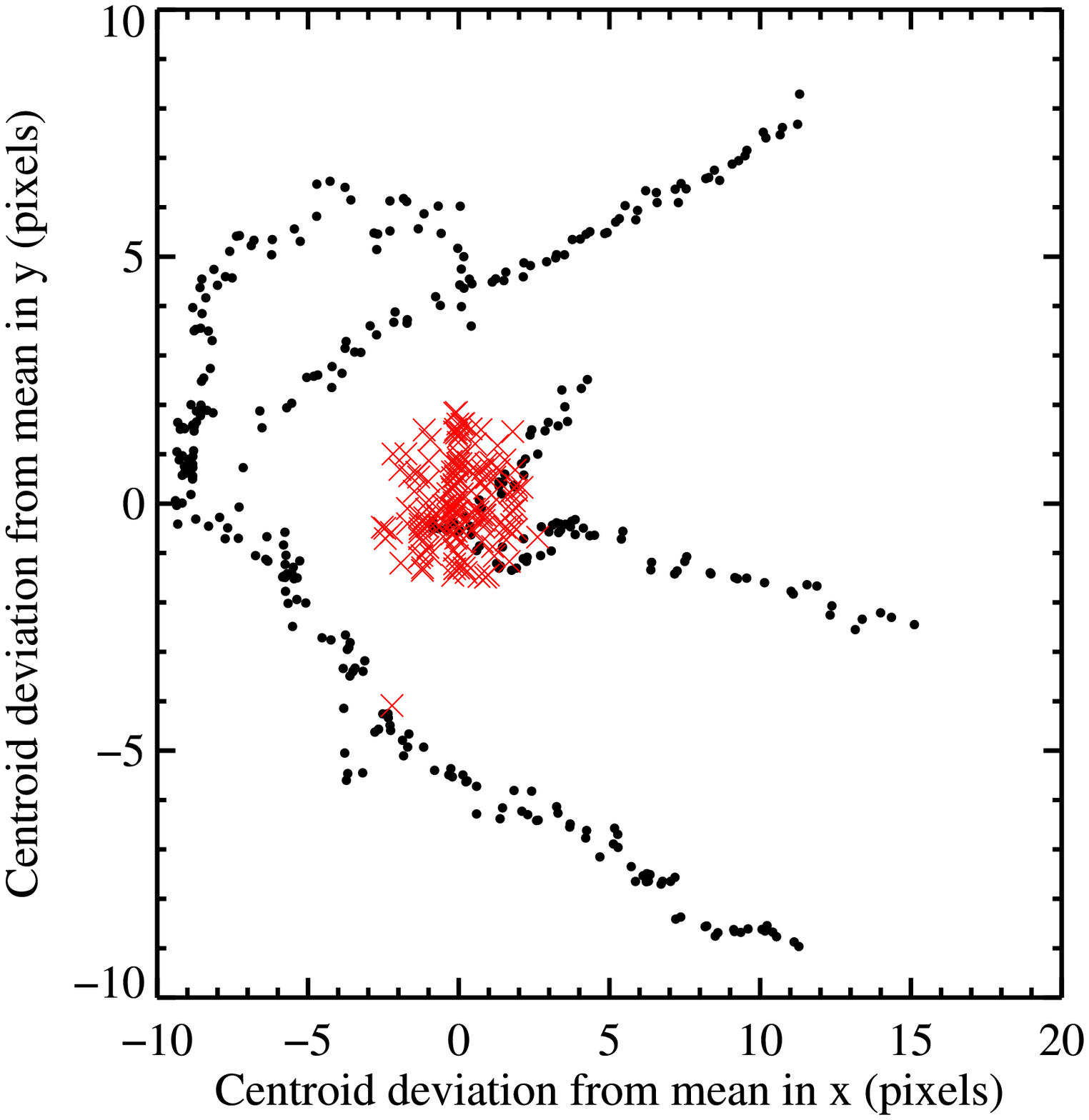}
\caption{\small{Position variations during a night with no specified maximum pointing error (black circles) and a night with maximum pointing error set to 3$''$ (crosses).}}
\label{fig:trackingcomp}
\end{center}
\end{figure}

For a given observation or group of observations in a night, ACP will first typically initiate an autofocus via FocusMax, which searches for mag 4-7 stars within a $2^{\circ}\times2^{\circ}$ region to focus with. Once focused, ACP starts a 4s `pointing exposure', which it subsequently plate-solves via PinPoint\footnote{http://pinpoint.dc3.com/} using the GSC \citep{lasker1996} to determine the pointing error. A corrective slew is applied, and the image sequence commences. Each full exposure, once read-out, is plate-solved once more to check the pointing. We apply a maximum pointing error of 3$''$, forcing a pointing update if a given image is determined to be misaligned by more than 3$''$. The benefits of doing so can be seen in Figure \ref{fig:trackingcomp}, where the centroid deviation is significantly reduced. This reduces the contribution to photometric noise from flat fielding errors; more on this in section \ref{sec:flatfield}.

The software tasks performed between exposures in a time-series run are: 

\begin{enumerate}
\item The previous image is read out and plate-solved, from which the astrometric residual, pointing error (magnitude and direction), mean FWHM, true focal length, true image centre, and camera sky position angle are calculated.
\item The centering is tested: if within the maximum pointing tolerance, no slew update is performed. If outside the maximum error tolerance, the auto-guiding is temporarily stopped, and a corrective slew is performed.
\item A guide star is selected from the guider image, with the star providing a signal-to-noise $\ge$ 3 for the shortest exposure interval chosen. The exposure is started.

\end{enumerate}

We typically see astrometric residuals from the plate-solve of $\sim$0.2$''$. The total inter-frame overhead between exposure end and exposure start is $\sim$ 15s.  Each frame takes $\sim$ 2.5s to read-out and download, meaning the overhead is dominated by the time taken to find the plate solution. The 15s quoted refers to the case where no subsequent pointing update is required after the plate-solve. When a pointing update is required (pointing error \textgreater 3$''$),  1-2s is added to the overhead time.

ACP also provides a web control interface, which made it an obvious choice for implementation in a distance-learning, group-based education scenario. When used by researchers, direct access to the PIRATE control PC is achieved by remote access of the control PC's desktop. Students, however, are not given direct access to the main control PC, and instead interact with the hardware through a proprietary server/client set-up that limits their control. The server-side application (`Switch Server') provides access to relay switches which are used to turn components on individually, and also has a suite of extra capabilities. These include: preventing the dome from being opened in daylight hours; closing the dome in case of excessively high humidity or strong winds; automatically cooling the two cameras in a monitored, staggered sequence; and checking the availability of the local internet connection, forcing a dome shut-down in case of lost internet connection. Further information on the operation of PIRATE's Switch Server can be found in \cite{lucas2010}.


\section{Data Acquisition, Reduction \& Commissioning Issues}

\subsection{Observing Strategy}
On a given night, the user cools the main imaging camera down automatically using the Switch Server to a target temperature of $\sim$ 40$^{\circ}$C below the current ambient temperature. Consequently the operating temperature of the CCD varies seasonally, usually between -10$^{\circ}$C in the Summer and -20$^{\circ}$C in the Winter. This ensures the camera is always operating at the lowest available temperature. Once the target temperature is achieved, ACP is started and the dome opened around sunset. ACP's automatic sky flat procedure is then initiated. This procedure automatically scales exposure times to achieve a target peak count of $\sim$ 33k ADU in each filter's flat fields. During the flat procedure, sidereal tracking of the mount does not occur. The flat fields are dithered so that any stellar sources present in the sky flats can be removed upon combination of each flat into the master flat field. Once the flat fields are obtained, the dome is then shut whilst waiting for the end of Nautical twilight (ACP allows observations through Astronomical twilight). During this time, the user takes calibration frames. 50 bias frames are taken first, to assist in completely flushing the chip of any residual charge from the flat fields. Then dark frames are taken in batches of 10 for each anticipated exposure time to be used for the science frames throughout the night. Typically these exposure times are 30, 45, 60, 90, and 120s. The dome is then reopened in time for the end of nautical twilight, and the observing schedule for the night commences.

\subsection{Photometric Reduction, Performance, and Light Curve Generation}

The data is available in real-time via direct FTP to the control PC, and users with direct access to the control PC may also inspect each image `live' after being read-out from the camera. Each morning the previous night's science and calibration frames are then transferred from the OAM to a local archive at the OU; the user may obtain the frames the following day. Reduction (in the research case) is performed by a custom IDL script that operates interactively, performing file renaming and sorting tasks and determining the presence of calibration frames, subsequently compiling IRAF \footnote{IRAF is distributed by the National Optical Astronomy Observatory, which is operated by the Association of Universities for Research in Astronomy (AURA) under cooperative agreement with the National Science 
Foundation.} scripts that make use of the DAOPHOT package \citep{stetson}. The DAOPHOT output is then fed back into an IDL script that performs differential photometry and generates the final light curves for each object in the frame. When used for teaching, the students do not make use of such scripts, and must process their frames `manually' using MaxIm DL.

The initial IDL script first compiles a list of available light frames from the previous night, and then iteratively displays each one in turn, asking the user to reject or accept each frame. This allows for the rejection of frames with defects, such as a lack of useable point sources in the frame, trails, or excessive pointing error. The user then accepts or rejects a final list of rejected frames, producing a master file list on acceptance. Lists are generated for each calibration frame type; discriminating flat fields by filter. If flat fields matching the chosen filter of the light frames do not exist, no flat fielding is performed. An IRAF script is generated for the task CCDPROC, which performs standard CCD calibrations and then executes the requisite post-calibration rotation commands for pre-flip light frames. 

The IDL routine first proposes a master frame according to a simple statistical test performed on all frames. The master frame proposed will be the frame that exhibits the highest standard deviation from the quartile of frames that have the lowest mean values. The mean test ensures a low sky background (high sky backgrounds are often indicative of Cirrus scattering), whilst selecting a frame with high standard deviation ensures many strong point source fluxes well above the background. The user can over-ride the proposed frame selection and select their own frame as the master frame. An alternative method would be to determine the frame with the minimum average FWHM, and designate this the master frame due to the superior seeing. This is not currently implemented.

Once defined, we use DAOFIND on the master frame (making use of a user-given average frame FWHM) to find all sources above a $6\sigma$ threshold. An initial PHOT task run on the master frame determines the master coordinate list, with only sources that do not suffer a PHOT task failure flag going into the master coordinate list. A linear transformation is then deduced between the master frame coordinates and those of all other frames, and this is used to generate the final co-ordinate lists for each frame. Final photometry is standardly performed with an aperture $3\times$ the input (master frame) FWHM, though the aperture size can be changed if the data warrants (e.g. in the case of close companions to the target that lie within or partially contaminate the measuring aperture).

Once the instrumental photometry is obtained, it is fed into another IDL script which places the whole night's photometry in memory as a data cube. Mid-exposure time in UTC then is converted to HJD for all frames.  We note the preference of using BJD over HJD, which can differ by as much as 4s \citep{eastman}, as well as calculating BJD from TT instead of UTC, which currently differ by 66.184s as of July 2011. We will be incorporating these changes into IDL scripts in the near future. The night's data is then investigated for PHOT task failures. Light curves that suffer a photometry failure in one or more frames are rejected from the cube. In case of bad frames still present in the data (which would lead to an excessive number of light curves being cut), we introduce a parameter (to be set by the user) defined as: $N_{star}/a$, where a frame is removed if it contains a fraction of photometry failures $> N_{star}/a$. Here $N_{star}$ is the number of stars in the frame, and $a$ is a free parameter. We find a value of $a=10$ to usually be sufficient. The cube is then sorted by magnitude, and an iterative ensemble compilation procedure begins. We employ a similar methodology to that of \cite{burke}, where for each star, all the corresponding comparison light curves are sorted according to a light curve `figure of merit'. Each comparison light curve is normalised to unity, and these light curves' RMS (i.e. the standard deviation about the mean) is used as the figure of merit. Starting with the best rated light curve (i.e. lowest RMS) for a given star, each subsequent light curve in the list is iteratively included in the ensemble, using inverse variance weights, where the weights derive from the IRAF photometric uncertainties. Note that the ensemble never contains the flux from the target star, as the target star is excluded from the list of light curves. The following refers to the production of the final light curve for one star only. The ensemble flux, $E_j$, for a given frame $j$, containing comparison stars $i=1...N$ is compiled as thus:

\begin{displaymath}
E_j=\frac{\sum_{i=1}^N \omega_{ij}F_{ij}}{\sum_{i=1}^N \omega_{ij}}
\end{displaymath}
Where $F_{ji}$ are the stars' individual fluxes, and the individual weights are given by:

\begin{displaymath}
\omega_{ij}=\frac{1}{\sigma_{ij}^2}
\end{displaymath}
The uncertainty in the ensemble is given by the standard deviation of the weighted average:

\begin{displaymath}
\sigma_{E_j} = \left [ \sum_{i=1}^N w_{ij} \right ]^{-1/2} 
\end{displaymath}

\begin{figure}[htp]
\begin{center}
\plotone{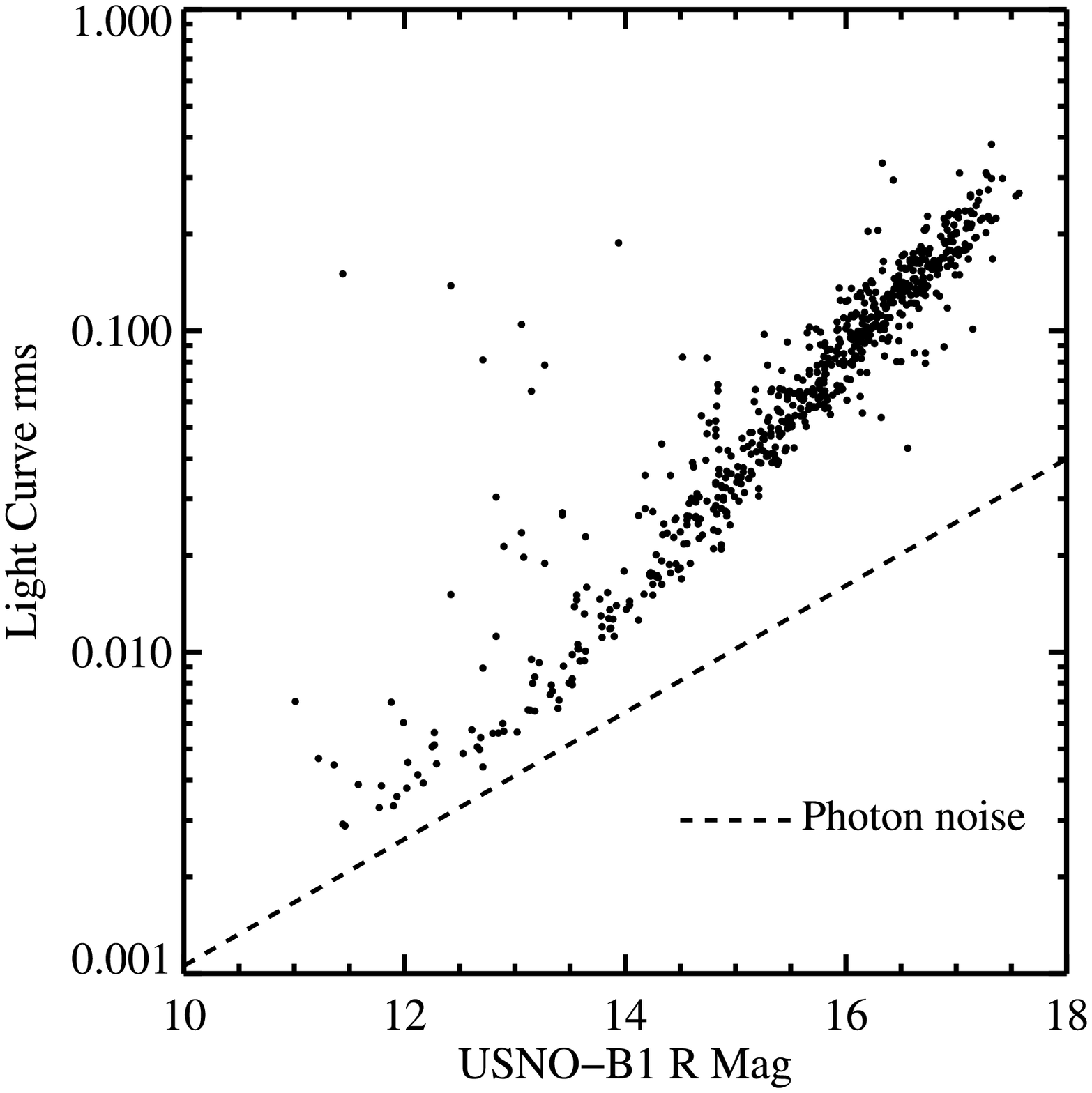}
\caption{\small{Light curve RMS deviation as a function of USNO-B1.0 R magnitude for the nights of 22/11/2009 and 23/11/2009, 60s exposures, WASP-12 field}}
\label{fig:rmsvmag}
\end{center}
\end{figure}

With each new star added to the ensemble, the differential target star light curve $d_* = (d_1\ldots d_N)$, where $d_{*,j} = \frac{F_{*,j}}{E_j}, j=1\ldots N$, is calculated from the recorded flux of the target star, $F_{*,j}$, and calculated ensemble flux, ${E_j}$. The light curve $d_*$ is then normalised by dividing through by its median value, and the RMS is computed. The final light curve, $d_{*,fin}$ makes use of the ensemble that produces the lowest RMS. Each star therefore has its own unique comparison ensemble. In Figure \ref{fig:rmsvmag}, we display the results of this method for the stars in the WASP-12 field on two consecutive nights with approximately 30\% \& 39\% moon illumination, at a separation of $\sim$ 142$^{\circ}$ and $\sim$ 132$^{\circ}$ from the moon respectively, comparing USNO-B1.0 \citep{monet} magnitudes with the RMS precision of PIRATE R band light curves from 60s exposures. The observation runs of 22/11/2009 and 23/11/2009 contain light curves of 759 and 730 objects, each consisting of 359 and 321 exposures respectively. We see from this plot that sub-percent precision is available for magnitudes $\sim$13.5 and less with an operating cadence of 76s (including the aforementioned overheads). We include for comparison the contribution to the uncertainty from photon noise alone.

\begin{figure}[htp]
\begin{center}
\plotone{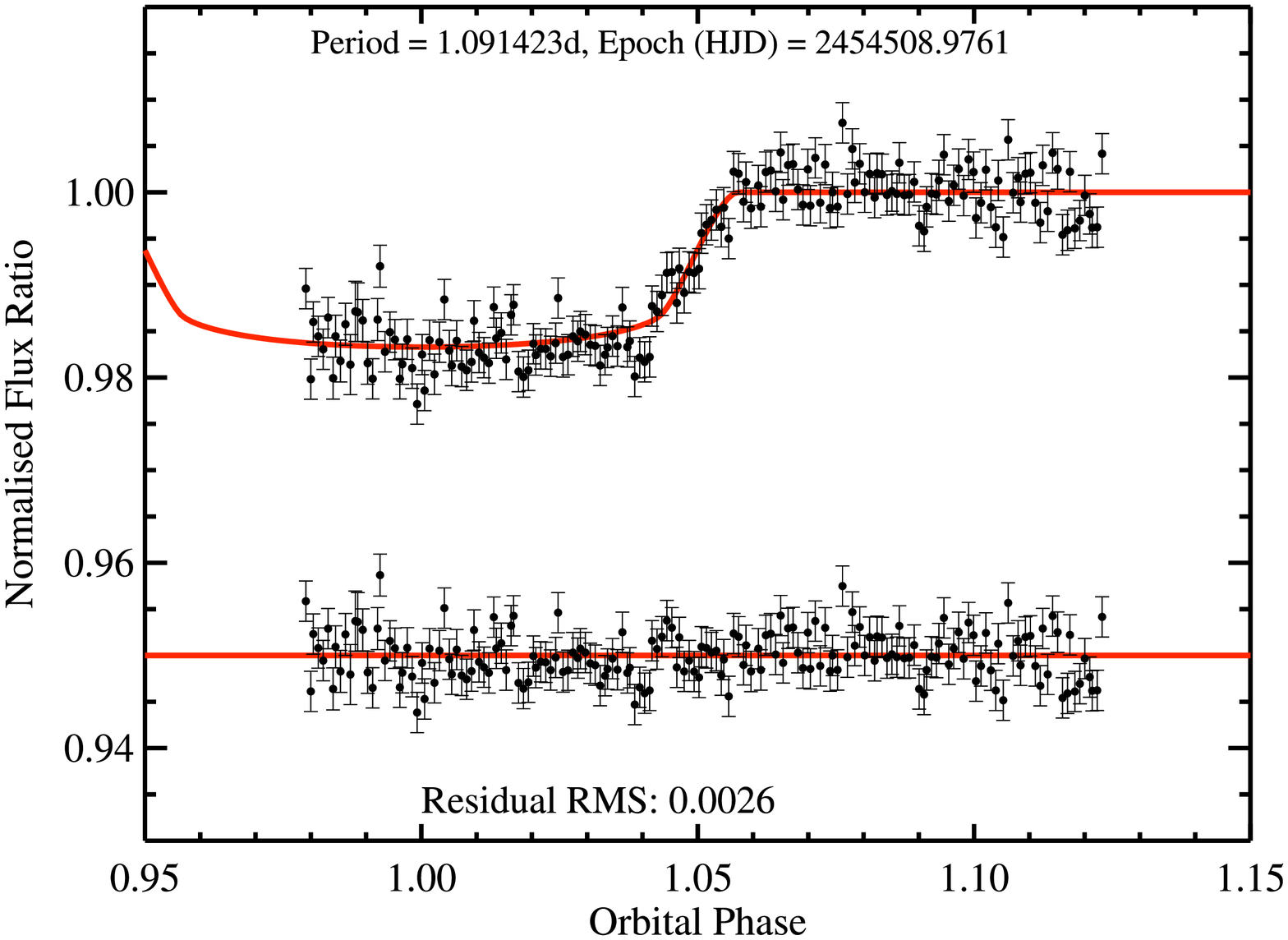}
\caption{\small{WASP-12b transit observation (R band) from 23/11/2009. Residuals from the model fit are shown offset by 0.05.}}
\label{fig:wasp12lc}
\end{center}
\end{figure}

As an example, this particular method of light curve compilation was used in the production of light curves of the transit of WASP-12b, for which 6 nights of data were captured during the months of October-November 2009, the details of which can be found in Haswell et al. (in preparation). We show the light curve from the night of 23/11/2009 in Figure \ref{fig:wasp12lc}, which has a post-fit residual RMS of $\sim$2.6 mmag. This data was used to confirm the published ephemeris for an HST visit investigating the exosphere of WASP-12b \citep{fossati}. The displayed model light curve is the published model light curve (fit to the data from the discovery paper) using the parameters of \cite{hebb}, and analytic light curve solutions of \cite{mandelagol}, using the limb-darkening coefficients of \cite{claret}.

%


We find that the entire light curve's RMS is an unsuitable figure of merit for determining how accurately a light curve has captured an object's inherent variability in certain cases. Substantial out of transit coverage of WASP-12 (out to phase 1.4 (not shown)) meant that the  RMS is an acceptable figure of merit for the observations shown in Fig. \ref{fig:wasp12lc}. Obviously, if one has prior knowledge of the transit ephemeris, as with these WASP-12 observations, solely the out-of-transit light curve should be used to calculate the figure of merit. In the case of the follow-up of candidate planets, or for other purposes (e.g. cluster surveys), the ephemeris might not be well constrained. The method's success is therefore subject to the amplitude, timing, and character of the star's intrinsic variability. 

When tackling variable stars with amplitudes over and above $\sim$10\%, the procedure will often select very faint, non-varying stars for inclusion within the ensemble, so long as their light curve RMS is less than the intrinsic variability of the target star. We showed in Fig. \ref{fig:eblc} the light curve of a larger amplitude ($\sim$20\%) variable, 1SWASPJ163420.90+424433.4 \citep{norton}, taken on 07/06/2010 by OU undergraduate students as part of their practical module. The data were reduced separately for inclusion here, with 4 comparison stars chosen by hand for combination through inverse-variance weighting into a comparison ensemble, as the RMS-minimisation method does not work for this target. We also find the procedure inadequate in some lower amplitude situations, such as a typical hot jupiter transit with insufficient out-of-transit coverage. For example, a light curve that captures 1 hour in-transit, egress, and 1 hour out-of-transit approximates a step function; and is bimodal in $d_{*}$, a scenario for which the RMS is a poor figure of merit.

To combat this, we are developing an extension of the technique that breaks each comparison light curve into multiple segments, and performs a linear regression in each segment, producing a selection of relevant statistics: segment RMS, residual segment RMS (after subtracting the fit), segment median value. By careful analysis of each light curve's segment statistics, as well the statistics of the segments as a whole (e.g. range or standard deviation of the segment medians), it will be possible to identify variable stars for exclusion from a comparison ensemble, whilst simultaneously allowing for any variable star to have an optimal comparison ensemble automatically determined. This technique will facilitate serendipitous PIRATE discoveries of previously unknown variable stars.

\subsection{Flat Field Inaccuracy and Optical Response Calibration}\label{sec:flatfield}
The systematic errors of flat fields as optical response calibrators are often overlooked in the application of differential photometry, as modern auto-guiding systems are sufficiently good at locking the stellar point-spread functions to fixed positions on the chip for the duration of a time series. As the observer is only interested in time-variability, systematic offsets in the magnitude zero-points of any two stars in the field can be ignored. PIRATE's GEM executes a `pier-flip' as the tracked target crosses the Meridian. This moves the OTA to the other side of the pier; and inverts the image of the stellar field with respect to pre-flip frames. This operation effectively moves each stellar point spread function to an entirely different part of the focal plane. To maintain continuity in the flux ratio between two stars across the pier flip, the vignetting must be determined perfectly. In reality, most flat fields suffer from inaccuracies at the $10^{-2}$ level \citep{manfroid}. We therefore expect and indeed see varying flux ratios between two objects in the field across the pier-flip. We term this effect a `Light Curve Discontinuity' (LCD).  Several mechanisms contribute to the flat field systematic errors:

\begin{itemize}

\item Uneven illumination from the source.  This can occur in sky flats, even when they image the sky null point \citep{chromey} as PIRATE's automatic procedure does. It will also occur in dome flats.

\item  Physical changes in the hardware between taking flat fields and science images that modify the response function of the system. In the case of PIRATE Mark 1, it is suspected that primary mirror `flop', which occurs despite the presence of mirror locks, and/or flexure in the micro-focuser assembly changes the vignetting function. It may also lead to the misalignment of the focal plane with the CCD surface, meaning consistent focus across the field of view is not achievable. In this scenario, the FWHM of the stellar profiles will be a function of chip position, and the measuring apertures used for photometry would see different levels of flux leakage due to the changed stellar profile after a pier-flip.

\item Scattered light in the optical tube assembly. Off-axis, unfocused stray light due to insufficient baffling can contribute to the flat field exposure. Scattered light is additive. If included in the flat field, this additive light is incorrectly used in the multiplicative flat field calibration. 
\end{itemize}

\begin{figure}[htp]
\begin{center}
\plotone{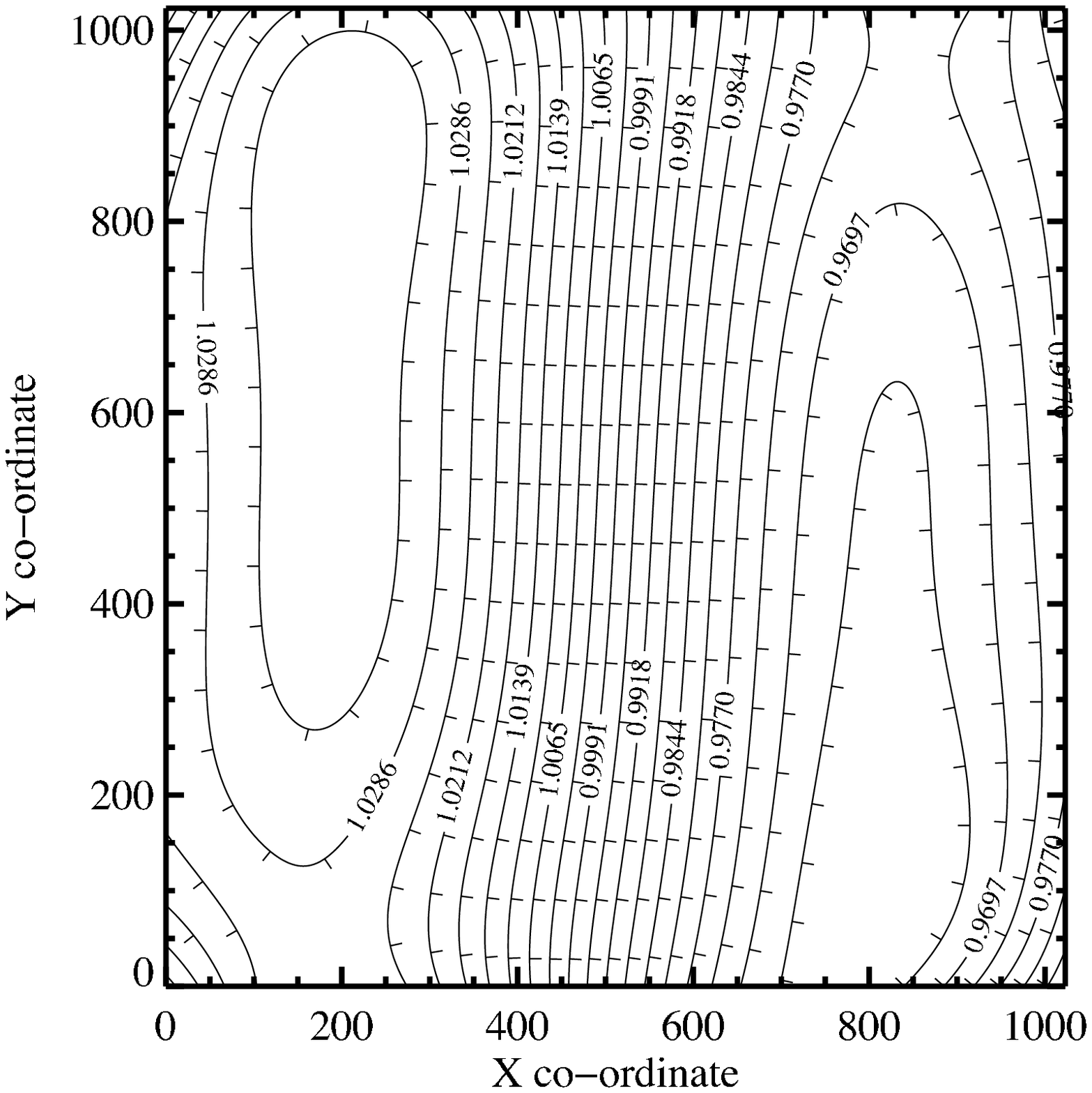}
\caption{\small{Sky background map produced by fitting to the sky background for all pre-flip frames and all post-flip frames, then taking the ratio of these two fits. The structure of this map suggests there may have been a fixed scattered light source with respect to the OTA.}}
\label{fig:lcdmap1}
\end{center}
\end{figure}

We have been unable to develop a wholly effective procedure for self-consistently calibrating the LCD effect without investing much of the night in taking calibration observations. We thoroughly explored these possible approaches:

We used data from the night of 23/07/2009, during which the moon had $3.2\%$ illumination, so we might expect sky background gradients to be limited. The data were processed in the usual manner, and investigated separately as two groups of `pre-flip' and `post-flip' frames. The frames in each group were median-combined, and a 6\textsuperscript{th} order Legendre polynomial was fit to the sky background in each of the two (pre and post) resultant frames using the IRAF task `IMSURFIT'. Note that the pre-flip frames were rotated 180$^{\circ}$. We denote the background fits by $A_{pre}(x,y)$ and $A_{post}(x,y)$, where $x$ and $y$ are image co-ordinates corresponding to positions on the sky (not pixel co-ordinates, due to the aforementioned rotation of the pre-flip frames). A simple ratio of the pre and post-flip background fits reveals any discrepancy in the sky background for a given star position. We denote this ratio as $A_{map}(x,y)=\frac{A_{pre}(x,y)}{A_{post}(x,y)}$. We show $A_{map}$ in Fig. \ref{fig:lcdmap1}. We note that it has structure predominantly in the $x$ direction, and displays a peak-to-peak variation of $\sim7\%$. The (non-differential) light curves of each of the $N$ stars (here $N=517$), which we denote by $F_i(t)$, where $i=1...N$ were median-combined to create an approximation (to first order) of the sky transparency, $\tilde{F}$. We assume each star remains at a fixed position in the image $(x_i,y_i)$ for the duration of the observing run. For all of the light curves (including the transparency function), the fluxes were averaged over time for pre-flip frames and also for post-flip frames, producing a single pre-flip and post-flip flux value for each light curve. We therefore use $F_{pre,i}$ and $F_{post,i}$ to refer to these time-averaged values. We define the LCD of a star to be:

\begin{displaymath}
\Delta F_i = \frac  {\sfrac   {F_{pre,i}} {F_{post,i}}}   {\sfrac {\tilde{F}_{pre}}{\tilde{F}_{post}}}
\end{displaymath}

\begin{figure}[htp]
\begin{center}
\plotone{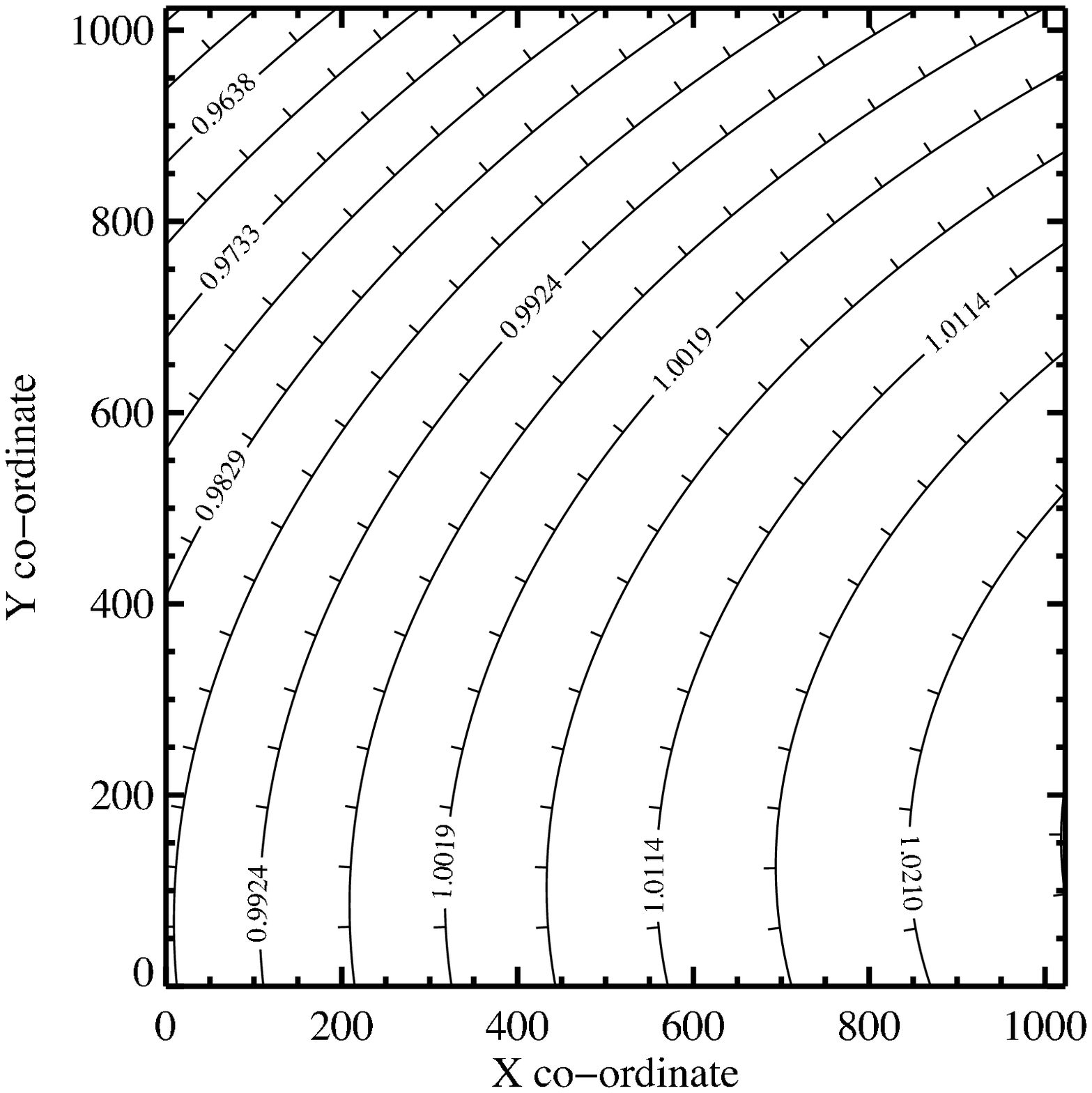}
\caption{\small{2\textsuperscript{nd} order polynomial fit to the flux deficits, $\Delta F_i$ }}
\label{fig:lcdfit1}
\end{center}
\end{figure}

We now have the ability to assess the applicability of the background `map'; if the shape of the sky background (after flat-fielding) provides an estimation of any residual error in the flat-fielding process, then it should correlate well with the observed LCDs. To compare the map with the LCDs, we fit a 2\textsuperscript{nd} order polynomial through least squares regression to the irregularly gridded $\Delta F_i$ values. The result of this fit can be seen in Fig. \ref{fig:lcdfit1}. As can be seen, far from correlating well, the pattern of the LCDs also exhibits greatest deviation in the $x$ direction, but with opposing orientation. The structure of the map in Fig. \ref{fig:lcdmap1} suggests a fixed and constant light source co-moving with the optical tube assembly, inducing the same background structure in both pre and post-flip frames, which is then amplified when the pre-flip frames are rotated. One would expect any transformation of the vignetting function to be recorded in the sky background, as this gives a continuous indication of the vignetting function. However, the presence of any scattered light in either the science frames or flat fields renders the true vignetting function for each side the pier flip unrecoverable. 

The scattered light structure present in the science frames could also have been present in the flat field, and thus the LCDs introduced to the data through the flat fielding process alone, but the un-flat fielded data shows the same LCD structure seen in Fig. \ref{fig:lcdfit1}. Given the dark night (one night after a new moon) for this data set, it is apparent that scattered (additive) light is likely always present, and greatly reduces the accuracy of standard flat-fielding and sky-flat procedures; the GEM simply highlights this inaccuracy.

\begin{figure}[htp]
\begin{center}
\plotone{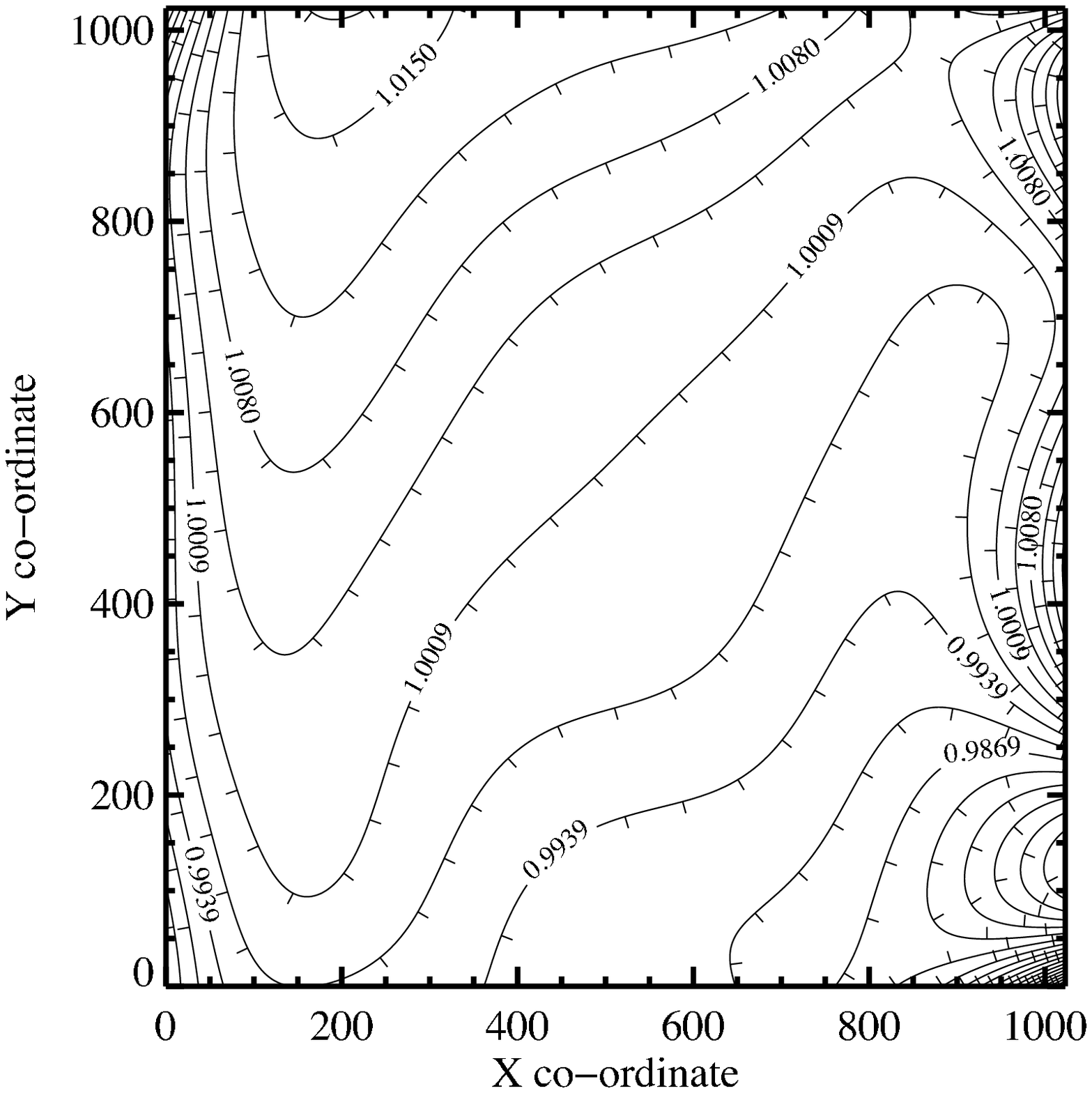}
\caption{\small{As for Fig. \ref{fig:lcdmap1}, but for the night of 23/11/2009}}
\label{fig:lcdmap2}
\end{center}
\end{figure}

\begin{figure}[htp]
\begin{center}
\plotone{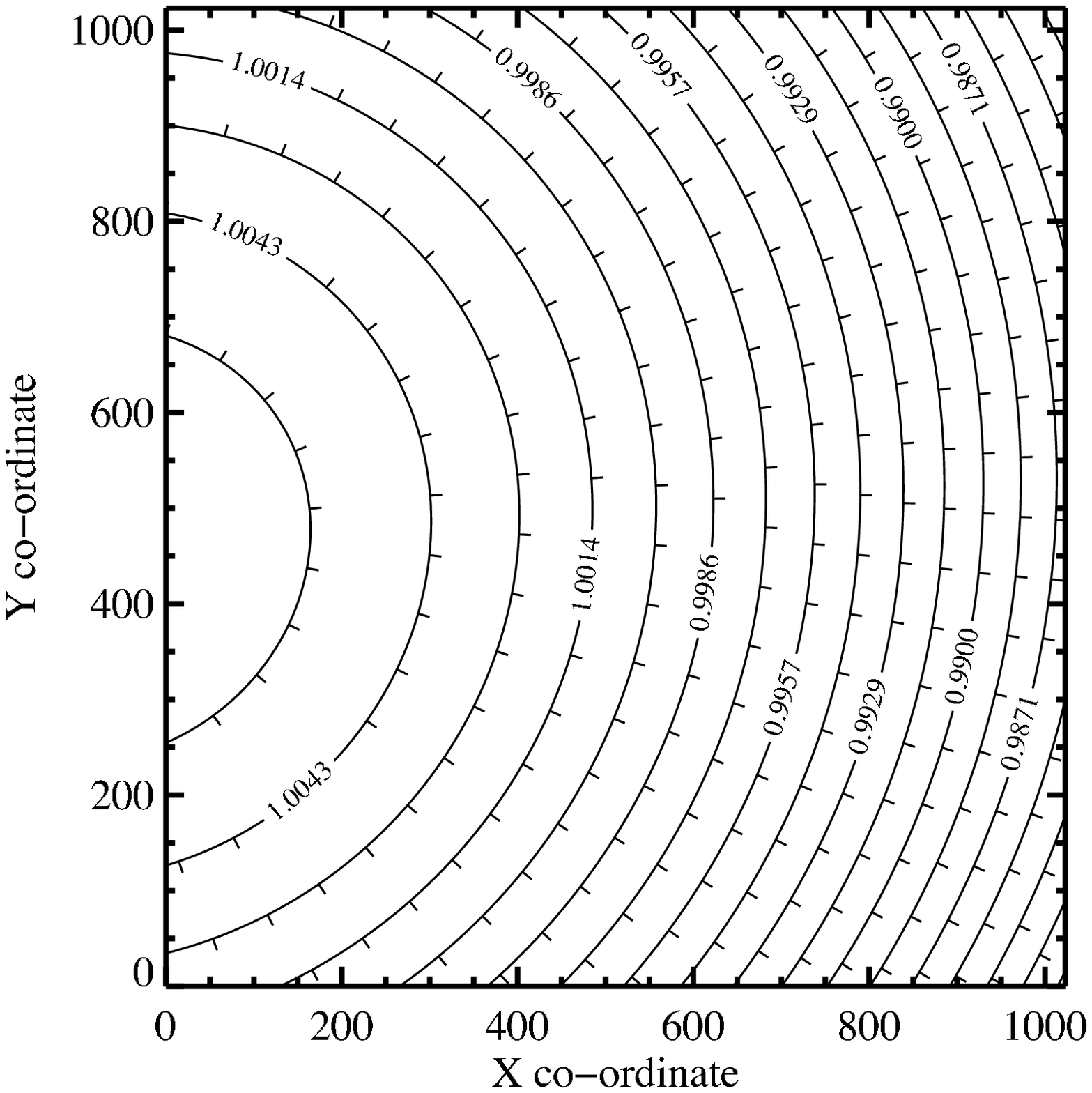}
\caption{\small{As for Fig. \ref{fig:lcdfit1}, but for the night of 23/11/2009}}
\label{fig:lcdfit2}
\end{center}
\end{figure}

The data of 23/07/2009 exhibit strong structure in the LCD map, and is hence a good demonstration of the problem. A more typical structure can be seen in Figs. \ref{fig:lcdmap2} and \ref{fig:lcdfit2}, from the night of 23/11/2009, WASP-12 field. The polynomial fit to the LCDs exhibits a smaller peak-to-peak amplitude of $\sim$2\%, and there is significant scatter in the residuals of this fit to $\Delta F_i$.

 We conclude that background fitting, sky flats, and even flat fielding are all ineffective at negating the percent-level LCD effect seen in PIRATE Mk. 1.  Instead, there are two routes to achieving correct photometric calibration across the pier-flip.

 The first of these involves creating a photometric super-flat. This is discussed in detail in \cite{boyle}, \cite{grauer}, \cite{manfroid}, \cite{regnault}, and \cite{selman}. At its most simple, this involves observing a set of standard stars at different positions in the field of view, and determining the position-dependent response to the standards in order to build the large-scale, low-frequency optical response of the system into the final flat field. Sky flats or dome flats are observed in tandem, with polynomial fits to their large scale variation made in order to `flatten' them. This retains the small scale, high frequency intra-pixel response, which can then be combined with the standard star-determined low frequency response function to create a `photometric super-flat'. In constructing the large-scale component, it is preferable to make multiple dithered observations of a cluster containing many standards that span the field of view in order to reduce the number of frames needed. As we expect the response to be different from one side of the pier to another, this involves observing the same cluster twice, once in the eastern sky, and again later in the western sky.

The second method involves lending a degree of freedom to the normalisation of each light curve section (pre \& post-flip), shifting the flux levels up and down to minimize the $\chi^{2}$ of a model fit. This method has its limitations in the extent of fore-knowledge required to work successfully. In the case of SuperWASP follow-up, the Markov-Chain Monte Carlo fitting routine of \cite{colliercameron2007} can take the light curves from each side of the pier as separate input light curves and include the normalisation factor as a free parameter in the fitting procedure. However, this algorithm works with `prior knowledge' of the light curve, in that its initial parameters from which the MCMC routine iterates are taken from the SuperWASP light curves. This prior expectation of the light curve's model parameters (including, crucially, the transit ephemeris) assists with the cross-flip normalisation. If there is no prior knowledge of the light curve, adjusting flux levels is almost certainly perilous. Take, for example, the worst case scenario of a pier-flip occurring mid-way through ingress or egress. In such a scenario it would prove difficult to determine if the flux deficit across the pier-flip is instrumental or astrophysical. Great caution is called for in interpreting such light curves.

\section{Summary and discussion}

PIRATE is a remotely-operable telescope facility built primarily from readily available `off-the-shelf' components, used in both education and research. In its education role, it has successfully supported small groups of simultaneous student users in their efforts to observe and classify variable stars in the SuperWASP archive. We showed that relatively inexpensive equipment can play a useful role in modern astrophysical research, including high-precision time-series photometry required for transiting Jupiter-size exoplanets. The PIRATE system employs a GEM which provides exceptional stability and all-sky pointing accuracy for its cost. The pier-flip introduces LCDs which are difficult to correct for by calibration procedures alone; the best strategy is to treat pre- and post-flip light curves as separate observing runs.

The OTA and camera combination described here awaits re-assembly in a neighbouring dome at the OAM. We note that the combination of a Paramount ME and the affordable Celestron C14 can be found in many installations world-wide.

The use of a remotely-operable telescope in the OU module S382 \footnote{http://www3.open.ac.uk/study/undergraduate/course/s382.htm} proved the feasibility of deploying such complex hardware in real-time to create an inspirational teaching tool for distance education. The great enthusiasm of the students involved, and the quality of the acquired data and of the scientific reports generated at the end of the project demonstrate the success of the PIRATE teaching project.  Use by level 2 (year 2) students is being implemented, and use by level 1 (year 1) OU students is currently under consideration. 

PIRATE has very recently undergone an OTA change to a PlaneWave Systems CDK17 f/6.8 Corrected Dall-Kirkham astrograph, which provides a bigger field of view and larger aperture (0.425m). Future work will involve linking ACP to the dome control via the ASCOM driver layer, allowing for full scripting of a night's events; adding a level of autonomy to the observing process. Use of ACP Scheduler\footnote{http://scheduler.dc3.com/} with the current weather station hardware and switch-server intelligence would effectively allow PIRATE to run in a fully autonomous, robotic mode, which will be explored as an operation mode for research purposes. The real-time remote operation will remain the only mode of operation for teaching applications, however, in order to make the learning process as interactive as possible.\\

We thank the referee for their constructive and insightful comments, which improved the paper. We are grateful for the continued support of Baader Planetarium towards optimising the dome for the PIRATE facility, and the on-site support provided by staff at the OAM. SH acknowledges the support of an STFC studentship.

\end{document}